\documentstyle[amssymb,twoside,psfig]{article}

\catcode`\@=11
\long\def\@makefntext#1{
\protect\noindent \hbox to 3.2pt {\hskip-.9pt
$^{{\eightrm\@thefnmark}}$\hfil}#1\hfill}               

\def\@makefnmark{\hbox to 0pt{$^{\@thefnmark}$\hss}}    

\def\ps@myheadings{\let\@mkboth\@gobbletwo
\def\@oddhead{\hbox{}
\rightmark\hfil\eightrm\thepage}
\def\@oddfoot{}\def\@evenhead{\eightrm\thepage\hfil
\leftmark\hbox{}}\def\@evenfoot{}
\def\sectionmark##1{}\def\subsectionmark##1{}}

\oddsidemargin=\evensidemargin
\newcounter{sectionc}\newcounter{subsectionc}\newcounter{subsubsectionc}
\renewcommand{\section}[1] {\vspace{12pt}\addtocounter{sectionc}{1}
\setcounter{subsectionc}{0}\setcounter{subsubsectionc}{0}\noindent
        {\tenbf\thesectionc. #1}\par\vspace{5pt}}
\renewcommand{\subsection}[1] {\vspace{12pt}\addtocounter{subsectionc}{1}
        \setcounter{subsubsectionc}{0}\noindent
        {\bf\thesectionc.\thesubsectionc. {\kern1pt \bfit #1}}\par\vspace{5pt}}
\renewcommand{\subsubsection}[1] {\vspace{12pt}\addtocounter{subsubsectionc}{1}
        \noindent{\tenrm\thesectionc.\thesubsectionc.\thesubsubsectionc.
        {\kern1pt \tenit #1}}\par\vspace{5pt}}
\newcommand{\nonumsection}[1] {\vspace{12pt}\noindent{\tenbf #1}
        \par\vspace{5pt}}

\newcounter{appendixc}
\newcounter{subappendixc}[appendixc]
\newcounter{subsubappendixc}[subappendixc]
\renewcommand{\thesubappendixc}{\Alph{appendixc}.\arabic{subappendixc}}
\renewcommand{\thesubsubappendixc}
        {\Alph{appendixc}.\arabic{subappendixc}.\arabic{subsubappendixc}}

\renewcommand{\appendix}[1] {\vspace{12pt}
        \refstepcounter{appendixc}
        \setcounter{figure}{0}
        \setcounter{table}{0}
        \setcounter{lemma}{0}
        \setcounter{theorem}{0}
        \setcounter{corollary}{0}
        \setcounter{definition}{0}
        \setcounter{equation}{0}
        \renewcommand{\thefigure}{\Alph{appendixc}.\arabic{figure}}
        \renewcommand{\thetable}{\Alph{appendixc}.\arabic{table}}
        \renewcommand{\theappendixc}{\Alph{appendixc}}
        \renewcommand{\thelemma}{\Alph{appendixc}.\arabic{lemma}}
        \renewcommand{\thetheorem}{\Alph{appendixc}.\arabic{theorem}}
        \renewcommand{\thedefinition}{\Alph{appendixc}.\arabic{definition}}
        \renewcommand{\thecorollary}{\Alph{appendixc}.\arabic{corollary}}
        \noindent{\tenbf Appendix \theappendixc #1}\par\vspace{5pt}}
\newcommand{\subappendix}[1] {\vspace{12pt}
        \refstepcounter{subappendixc}
        \noindent{\bf Appendix \thesubappendixc. {\kern1pt \bfit #1}}
        \par\vspace{5pt}}
\newcommand{\subsubappendix}[1] {\vspace{12pt}
        \refstepcounter{subsubappendixc}
        \noindent{\rm Appendix \thesubsubappendixc. {\kern1pt \tenit #1}}
        \par\vspace{5pt}}

\newcommand{\textlineskip}{\baselineskip=13pt}
\newcommand{\smalllineskip}{\baselineskip=10pt}

\def\eightcirc{
\begin{picture}(0,0)
\put(4.4,1.8){\circle{6.5}}
\end{picture}}
\def\eightcopyright{\eightcirc\kern2.7pt\hbox{\eightrm c}}

\newcommand{\copyrightheading}[1]
        {\vspace*{-2.5cm}\smalllineskip{\flushleft
        {\footnotesize International Journal of Modern Physics C #1}\\
        {\footnotesize $\copyright$\, World Scientific Publishing
         Company}\\
         }}

\newcommand{\publisher}[2]{{\begin{center}\footnotesize\smalllineskip
        Received #1\\
        Revised #2
        \end{center}
        }}

\def\abstracts#1#2#3{{
        \centering{\begin{minipage}{4.5in}\footnotesize\baselineskip=10pt
        \parindent=0pt #1\par
        \parindent=15pt #2\par
        \parindent=15pt #3
        \end{minipage}}\par}}

\def\keywords#1{{
        \centering{\begin{minipage}{4.5in}\footnotesize\baselineskip=10pt
        {\footnotesize\it Keywords}\/: #1
        \end{minipage}}\par}}

\newcommand{\bibit}{\nineit}

\renewenvironment{thebibliography}[1]
        {\frenchspacing
         \ninerm\baselineskip=11pt
         \begin{list}{\arabic{enumi}.}
        {\usecounter{enumi}\setlength{\parsep}{0pt}
         \setlength{\leftmargin 12.7pt}{\rightmargin 0pt} 
         \setlength{\itemsep}{0pt} \settowidth
        {\labelwidth}{#1.}\sloppy}}{\end{list}}

\newcounter{itemlistc}
\newcounter{romanlistc}
\newcounter{alphlistc}
\newcounter{arabiclistc}

\newcommand{\fcaption}[1]{
        \refstepcounter{figure}
        \setbox\@tempboxa = \hbox{\footnotesize Fig.~\thefigure. #1}
        \ifdim \wd\@tempboxa > 5in
           {\begin{center}
        \parbox{5in}{\footnotesize\smalllineskip Fig.~\thefigure. #1}
            \end{center}}
        \else
             {\begin{center}
             {\footnotesize Fig.~\thefigure. #1}
              \end{center}}
        \fi}

\newcommand{\tcaption}[1]{
        \refstepcounter{table}
        \setbox\@tempboxa = \hbox{\footnotesize Table~\thetable. #1}
        \ifdim \wd\@tempboxa > 5in
           {\begin{center}
        \parbox{5in}{\footnotesize\smalllineskip Table~\thetable. #1}
            \end{center}}
        \else
             {\begin{center}
             {\footnotesize Table~\thetable. #1}
              \end{center}}
        \fi}

\def\@citex[#1]#2{\if@filesw\immediate\write\@auxout
        {\string\citation{#2}}\fi
\def\@citea{}\@cite{\@for\@citeb:=#2\do
        {\@citea\def\@citea{,}\@ifundefined
        {b@\@citeb}{{\bf ?}\@warning
        {Citation `\@citeb' on page \thepage \space undefined}}
        {\csname b@\@citeb\endcsname}}}{#1}}

\newif\if@cghi
\def\cite{\@cghitrue\@ifnextchar [{\@tempswatrue
        \@citex}{\@tempswafalse\@citex[]}}
\def\citelow{\@cghifalse\@ifnextchar [{\@tempswatrue
        \@citex}{\@tempswafalse\@citex[]}}
\def\@cite#1#2{{$\null^{#1}$\if@tempswa\typeout
        {IJCGA warning: optional citation argument
        ignored: `#2'} \fi}}

\def\pmb#1{\setbox0=\hbox{#1}
        \kern-.025em\copy0\kern-\wd0
        \kern.05em\copy0\kern-\wd0
        \kern-.025em\raise.0433em\box0}

\def\fnt#1#2{\footnotetext{\kern-.3em
        {$^{\mbox{\scriptsize #1}}$}{#2}}}

\def\ps@myheadings{%
    \let\@oddfoot\@empty\let\@evenfoot\@empty
    \def\@evenhead{\slshape\leftmark\hfil}
    \def\@oddhead{\hfil{\slshape\rightmark}}
    \let\@mkboth\@gobbletwo
    \let\sectionmark\@gobble
    \let\subsectionmark\@gobble
    }
\font\tenrm=cmr10
\font\tenit=cmti10
\font\tenbf=cmbx10
\font\bfit=cmbxti10 at 10pt
\font\ninerm=cmr9
\font\nineit=cmti9

\font\eightrm=cmr8

\def\qed{\hbox{${\vcenter{\vbox{                    
   \hrule height 0.4pt\hbox{\vrule width 0.4pt height 6pt
   \kern5pt\vrule width 0.4pt}\hrule height 0.4pt}}}$}}

\def\bsc{{\sc a\kern-6.4pt\sc a\kern-6.4pt\sc a}}       
\def\bflatex{\bf L\kern-.30em\raise.3ex\hbox{\bsc}\kern-.14em
T\kern-.1667em\lower.7ex\hbox{E}\kern-.125em X}

\begin{document}
\setlength{\textheight}{7.7truein}  

\thispagestyle{empty}

\markboth{\protect{\footnotesize\it T. P\"oschel et al.}}{\protect{\footnotesize\it 
 VIOLATION OF MOLECULAR CHAOS IN DISSIPATIVE GASES}}

\normalsize\textlineskip

\setcounter{page}{1}

\copyrightheading{}                     

\vspace*{0.88truein}

\centerline{\bf VIOLATION OF MOLECULAR CHAOS IN DISSIPATIVE GASES}\vspace*{0.05truein}

\vspace*{0.37truein}
\centerline{\footnotesize THORSTEN P\"OSCHEL, NIKOLAI V. BRILLIANTOV, 
and THOMAS SCHWAGER}
\baselineskip=12pt
\centerline{\footnotesize\it Humboldt-Universit\"at zu Berlin, Charit\'e, Institut f\"ur Biochemie, }
\baselineskip=10pt
\centerline{\footnotesize\it Monbijoustra{\ss}e 2, D-10117 Berlin, Germany}
\centerline{\footnotesize\it thorsten.poeschel@charite.de, nikolai.brilliantov@charite.de, thomas.schwager@charite.de}

\vspace*{0.225truein}
\publisher{(received date)}{(revised date)}

\vspace*{0.25truein} 
\abstracts{Numerical simulations of a dissipative
  hard sphere gas reveal a dependence of the cooling rate on
  correlation of the particle velocities due to inelastic collisions. We
  propose a coefficient which characterizes the velocity correlations
  in the two-particle velocity distribution function and express the
  temperature decay rate in terms of this coefficient. The analytical
  results are compared with numerics.}{}{}

\vspace*{5pt} 
\keywords{kinetic gas theory, many-particle systems,
  dissipative gases, granular gases, velocity correlations}

\vspace*{1pt}\textlineskip
\section{Introduction}
\vspace*{-0.5pt}

\noindent The Boltzmann equation for the velocity distribution 
$f\left(\vec{v},t\right)$ of particles of a gas has been derived with
the use of the {\em molecular chaos} assumption, which implies that
all correlations between the velocities of particles may be
neglected\cite{Boltzmann:1896}. This equation was also a starting
point to build up a kinetic theory of dissipative gases where the
particles collide inelastically (for an overview see, e.g.\cite{LNP}).
Whereas the molecular chaos assumption is justified for dilute
molecular gases, there are two reasons for its violation for the case
of moderately dense dissipative gases: First the positions of the
particles are not independent due to excluded volume effects which
gives rise to the Enskog factor.\cite{ChapmanCowling:1970} A second
source of correlations is the dissipative nature of particle
interaction. Since dissipation concerns only the normal component of
the relative velocity of colliding particles, the angle between their
velocity vectors after the collision is always smaller than the angle
just before the collision. The latter effect leads to aligning of the
trajectories of adjacent particles which contradicts the molecular
chaos assumption. Obviously this effect increases with increasing
inelasticity, expressed by the coefficient of restitution $\epsilon$
which relates the normal component of the relative particle velocity
after and before a collision. This expectation has been confirmed by
studies of randomly driven Granular Gases\cite{Pagonabaretal:2002},
where it was shown that the molecular chaos assumption becomes less
accurate with increasing degree of particle inelasticity. Instead, the
velocities of the colliding pair are noticeably correlated; this
effect depends on the restitution coefficient and on the gas
density.\cite{SotoMareschal:2001} On the macroscopic (hydrodynamic)
scale these velocity correlations cause vortex structures in the
velocity field of a spatially homogeneous dissipative
gas.\cite{BritoErnst:1998}

In spite that obviously the microscopic correlations of the velocities
of adjacent particles play an important role for the kinetics of
dissipative gases, the particular form of the multi-particle velocity
distribution function, which describes these correlation is not known.
It is the aim of this paper to present the first (to our knowledge)
attempt to construct the two-particle correlation function of a simple
dissipative gas.

\section{Boltzmann-Enskog equation for hard spheres}
\vspace*{-0.5pt}
\noindent
In kinetic gas theory the evolution of the velocity distribution
function $f\left(v,t\right)$ is governed by the Boltzmann
equation\cite{Boltzmann:1896}
\begin{equation}
  \label{eq:BE}
  \frac{\partial}{\partial t}f\left(\vec{v}_1,t\right) = I(f,f)
\end{equation}
with the collision integral
\begin{equation}
  \label{eq:CI}
I(f,f) \equiv \sigma^2 \int d \vec{v}_2 \int d\vec{e} \, 
\Theta\left(-\vec{v}_{12} \cdot \vec{e}\right) \left|\vec{v}_{12} 
\cdot \vec{e}\right| 
\left[\chi f\left(\vec{v}_1^{\prime\prime},t\right)
f\left(\vec{v}_2^{\prime\prime},t\right)-
f\left(\vec{v}_1,t\right)f\left(\vec{v}_2,t\right) \right]\,.
\end{equation}
The velocities $\vec{v}_1$ and $\vec{v}_2$ are the pre-collision
velocities of the {\em direct collision}, and the unit vector
$\vec{e}$ points from the center of particle 2 with velocity
$\vec{v}_2$ to the center of particle 1 with velocity $\vec{v}_1$.
Here $\vec{v}_{12}\equiv \vec{v}_1-\vec{v}_2$, $\sigma$ is the
particle diameter and $\Theta(x)$ is the Heaviside step-function. The
velocities $\vec{v}_1^{\prime\prime}$, $\vec{v}_2^{\prime\prime}$
correspond to pre-collision velocities of the {\em inverse collision},
which ends up after the collision with the velocities $\vec{v}_1$ and
$\vec{v}_2$.

For a gas of inelastically colliding particles the function 
\begin{equation}
\label{eq:CHI}
\chi\equiv \frac{\left|\vec{g}^{\prime\prime}\right|}{\left|\vec{g}\right|}
\frac{{\cal D}\left(\vec{v}^{\prime\prime}_1,\vec{v}^{\prime\prime}_2 \right)}
{{\cal D}\left(\vec{v}_1,\vec{v}_2 \right)} 
\end{equation} 
accounts for the Jacobian of the transition from the pre-collision
velocities $\vec{v}_1^{\prime\prime}$, $\vec{v}_2^{\prime\prime}$ of
the inverse collision to those of the direct collision $\vec{v}_1$,
$\vec{v}_2$ and for the ratio of the lengths of the collision
cylinders of the inverse and the direct collision. These are
proportional to the normal components of the relative velocities, i.e.
$\vec{g}\equiv \left(\vec{v_{12}}\cdot \vec{e}\right)\vec{e}$ for the
direct collision, with the analogous expression for the inverse
collision.  The ratio
\begin{equation}
  \label{eq:COR}
  \epsilon\equiv \left|\vec{g} \right|/\left|\vec{g}^{\prime\prime}\right|
\end{equation}
defines the coefficient of restitution $\epsilon$ which describes the
inelasticity of the particles. Using this coefficient the velocities
of two colliding particles before and after a collision can be
related. In particular, these relations read for the inverse collision
\begin{eqnarray}
  \label{eq:coll}
    \vec{v}^{\prime\prime}_1&=&\vec{v}_1 - \frac{1+\epsilon}{2\epsilon} 
\left(\vec{v}_{12}\cdot \vec{e}\right)\vec{e} \\
    \vec{v}^{\prime\prime}_2&=&\vec{v}_2 + \frac{1+\epsilon}{2\epsilon} 
\left(\vec{v}_{12}\cdot \vec{e}\right)\vec{e}  \, .
\end{eqnarray}
The coefficient of restitution is the central quantity in the theory
of granular gases. Experiments\cite{BridgesHatzesLin:1984} as well as
theory
\cite{BrilliantovSpahnHertzschPoeschel:1994,SchwagerPoeschel:1998,Ramirez:1999}
show clearly that the coefficient of restitution is not a constant but
depends on the impact velocity. Nevertheless, for the sake of
simplicity of calculation it is frequently assumed that this
coefficient is a material constant. With the assumption $\epsilon={\rm
  const.}$ from (\ref{eq:CHI}) we find $\chi=\epsilon^{-2}$.

\section{Correlations in dissipative gases}
\noindent The most important precondition for the derivation of 
(\ref{eq:BE},\ref{eq:CI}) is the assumption of molecular chaos
(Sto{\ss}zahlansatz) which expresses the two-particle distribution
function as a product of the one-particle distribution functions
\begin{equation}
  \label{eq:MolChaos}
  f_2\left(\vec{v}_1, \vec{v}_2,t\right)= 
f\left(\vec{v}_1, t\right)f\left(\vec{v}_2,t\right)\,
\end{equation}
which is identical with the assumption of vanishing correlations
between successive collisions of each particle. For gases of particles
of finite sizes this assumption is violated due to finite volume
effects (i.e. two particles can not occupy the same volume). This
gives rise to the Enskog factor\cite{resibua} $g_2(\sigma)$
\begin{equation}
  \label{eq:Enskog}
  f_2\left(\vec{v}_1, \vec{v}_2\right)\approx g_2(\sigma) 
f\left(\vec{v}_1, t\right)f\left(\vec{v}_2,t\right)\,
\end{equation}
which accounts for the increasing collision frequency in such gases.
This quantity equals the contact value of the pair correlation
function $g_2(\sigma)$. For systems of hard discs it
reads:\cite{HansenMcDinald1986}
\begin{equation}
\label{2DCarnahanStarling}
g_2(\sigma)=\frac{1-\frac{7}{16} \eta}{( 1- \eta )^2} \, , 
\end{equation}
where $\eta=\pi n \sigma^2/4$ is the packing fraction, $n=N/V$ is the
number density with $N$ being the number of particles in the system
and $V$ is the two-dimensional volume (area). The Enskog factor
obviously accounts only for static spatial correlations. Any dynamic
correlations between the velocities of particles are still ignored.

\begin{figure}[htbp]
  \vspace*{13pt}
  \centerline{\psfig{file=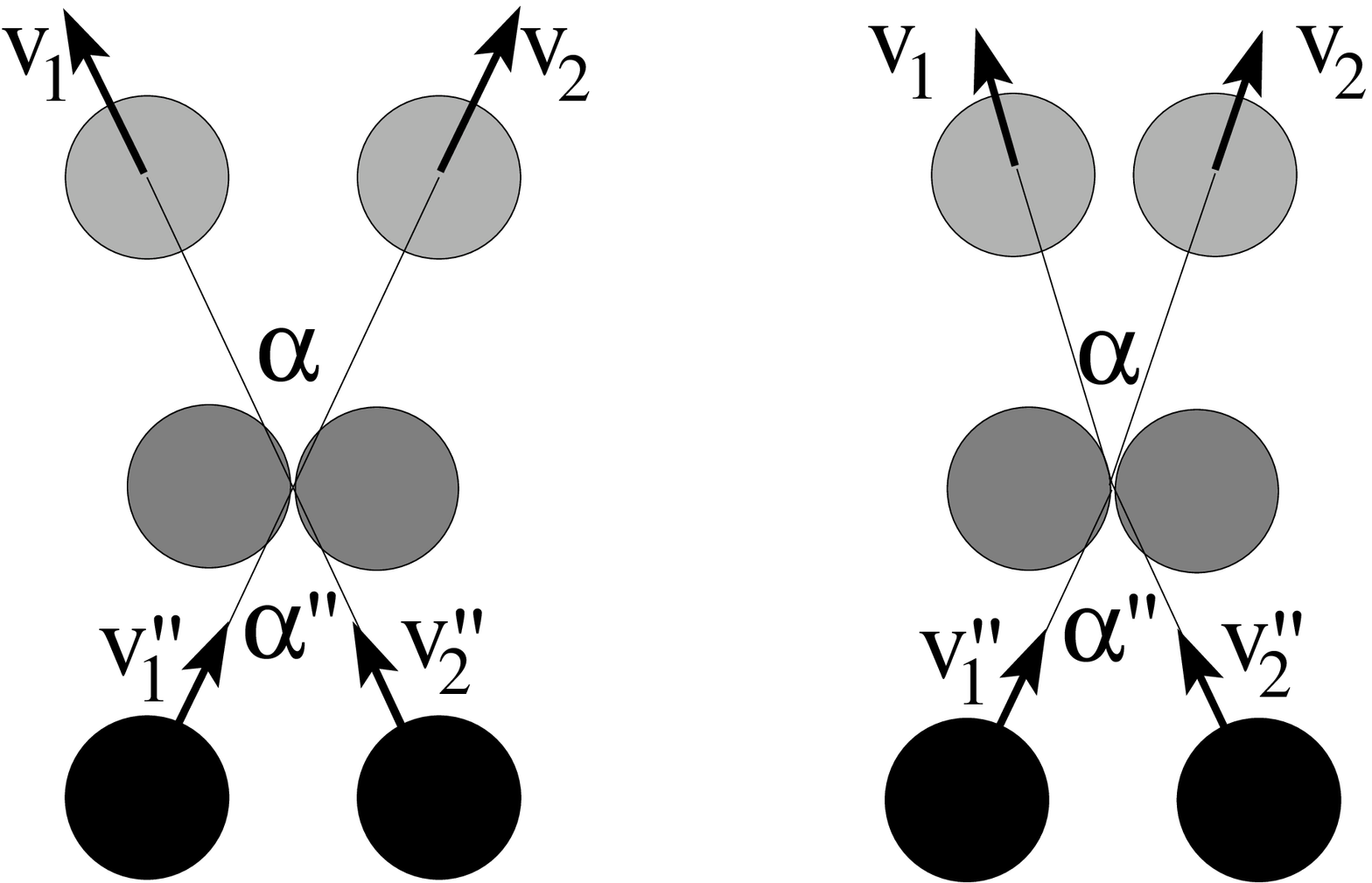,width=6cm}}
  \vspace*{13pt} \fcaption{Elastic (left) and inelastic (right)
    spheres before (black), during (dark grey) and after a collision
    (light grey). For elastic spheres the incoming angle equals the
    outgoing angle ($\alpha=\alpha^{\prime\prime}$) whereas for
    inelastic collisions $\alpha<\alpha^{\prime\prime}$.}
\label{fig:collision}
\end{figure}

For the case of dissipative gases there is an additional source of
correlations. These arise since the restitution coefficient affects
only the normal component of the relative velocity of a colliding
pair, leaving the tangential relative velocity unchanged. If two
inelastic particles collide with a (pre-collision) angle
$\alpha^{\prime\prime}$ the angle $\alpha$ after the collision will be
smaller, $\alpha< \alpha^{\prime\prime}$ (see Fig.
\ref{fig:collision}).

As a consequence subsequent collisions lead to more and more aligned
traces of neighboring particles. It has been
shown\cite{BritoErnst:1998,NoijeErnstBritoOrza:1997} that these
correlations give rise to convection cells in an otherwise homogeneous
granular gas, i.e., to macroscopic correlations in the velocity field
even in absence of spatial inhomogeneities.

Since the particles are aligned in macroscopic vortices, such velocity
correlations lead to a reduced collision frequency as compared with
the collision frequency of a gas of elastic particles of the same
temperature. The reduction of the collision frequency leads in its
turn to a reduction of the cooling rate, i.e., to a retarded decay of
temperature even in the stage of homogeneous cooling. Therefore, to
estimate the importance of velocity correlations we analyze the
cooling rate of a dissipative gas.

\section{Temperature decay of dissipative gases}
\noindent Assume a homogeneous dissipative gas of particles 
with velocities which obey a distribution function
$f\left(\vec{v}\right)$. For $\epsilon={\rm const.}$ the Boltzmann
equation decouples into a set of simpler
equations\cite{EsipovPoeschel:1995}, one for the time dependent
temperature
\begin{equation}
  \label{eq:TDef}
  T(t)\equiv \frac{m}{3n}\int d\vec{v} v^2 f\left(\vec{v},t\right) 
\end{equation}
and another one for the time independent (normalized) distribution
function. With the assumption that the velocities obey a Maxwell
distribution it has been shown by scaling arguments that temperature
decays due to dissipative collisions as
\begin{equation}
  \label{eq:Haff}
  T(t)=\frac{T_0}{\left(1+t/\tau_0\right)^2}
\end{equation}
with $T_0$ being the initial temperature and $\tau_0$ being a scaling
factor which will be discussed in detail below. Equation
(\ref{eq:Haff}) is known as Haff's law\cite{Haff:83}. Later deviations
from the Maxwell distribution have been
found\cite{NoijeErnst:1998,BrilliantovPoeschelStability:2000},
however, the temperature decay law is still correct.

\begin{figure}[htbp]
  \vspace*{13pt}
  \centerline{\psfig{file=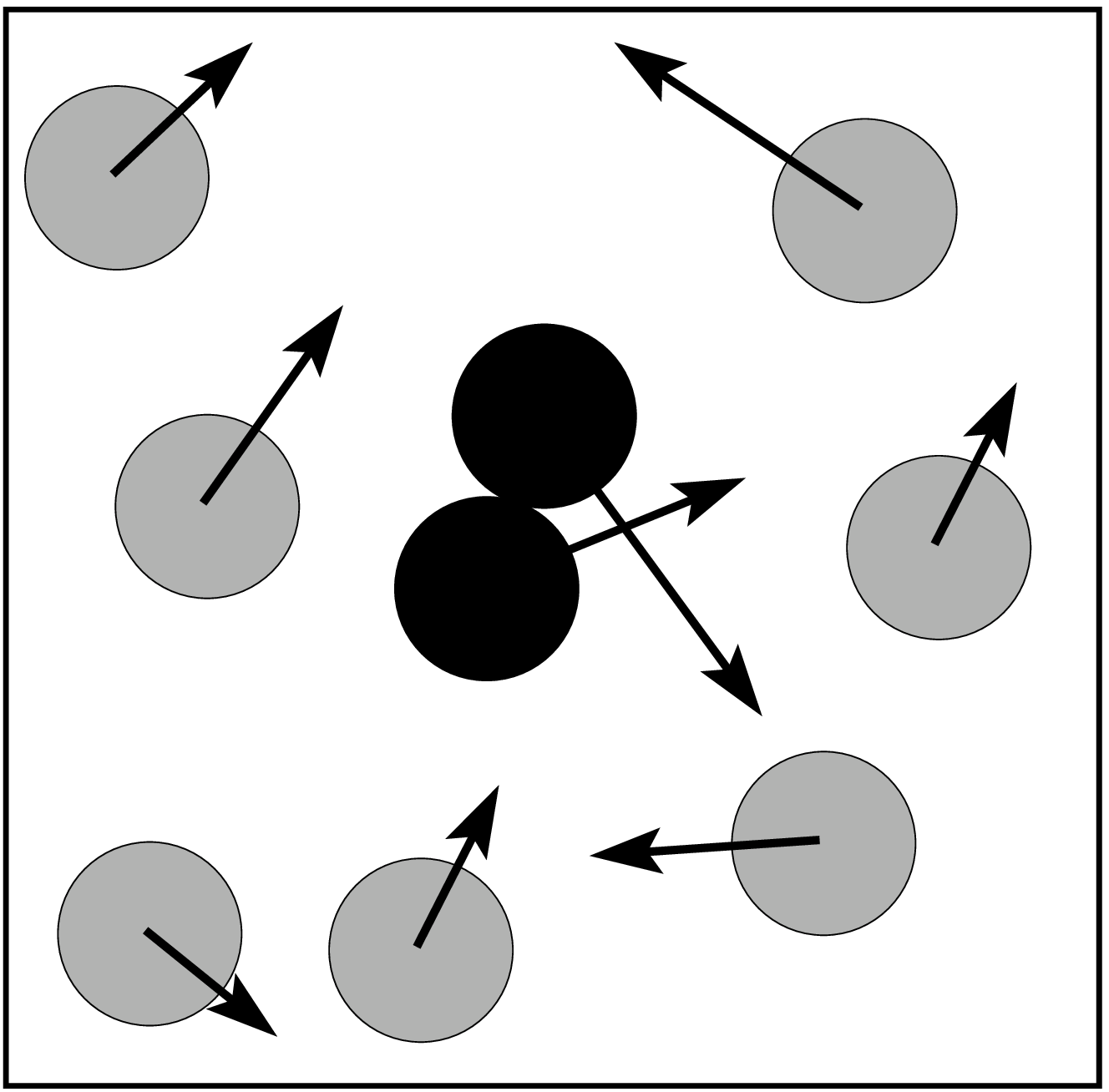,width=4cm}~~\psfig{file=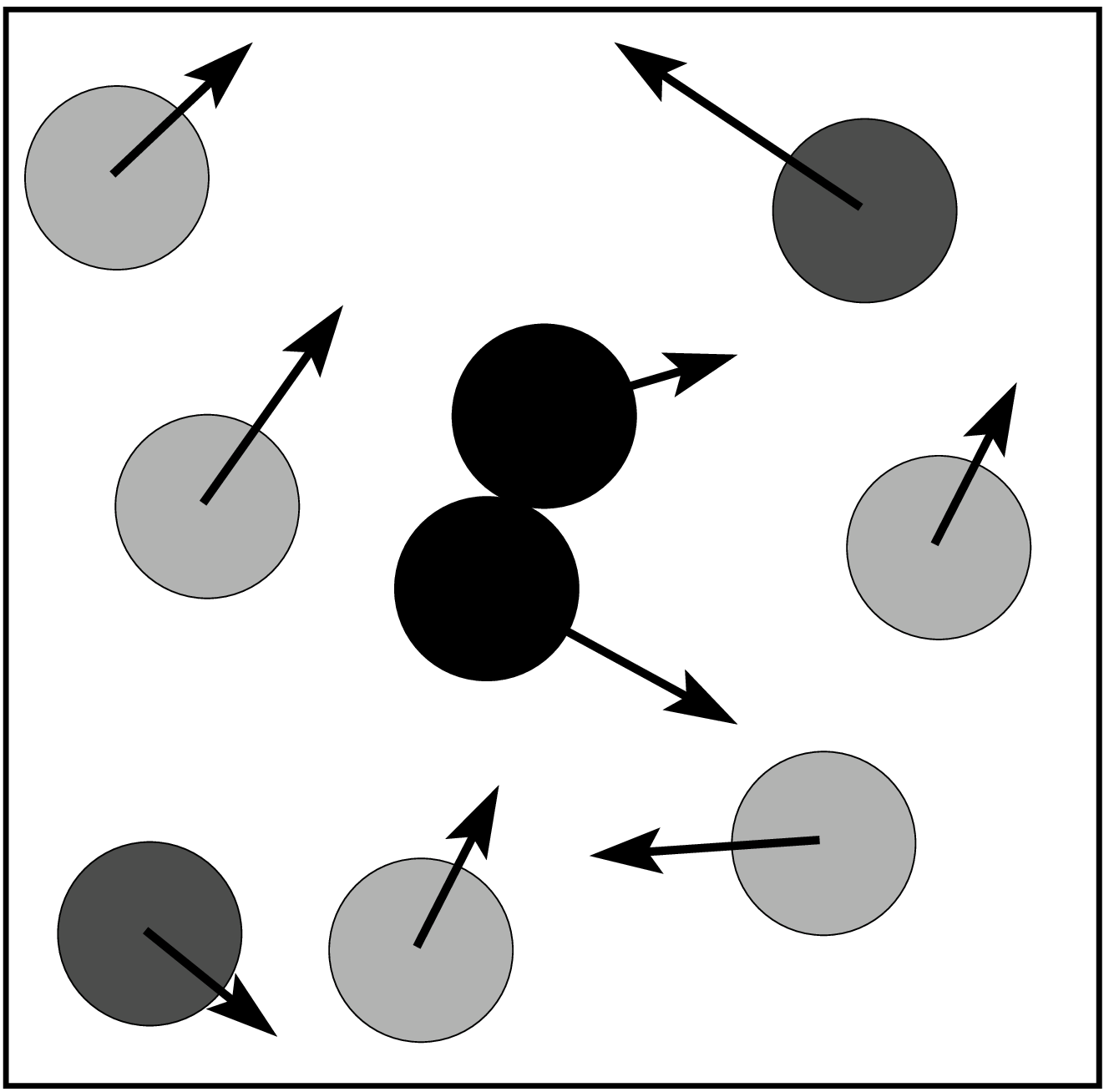,width=4cm}~~\psfig{file=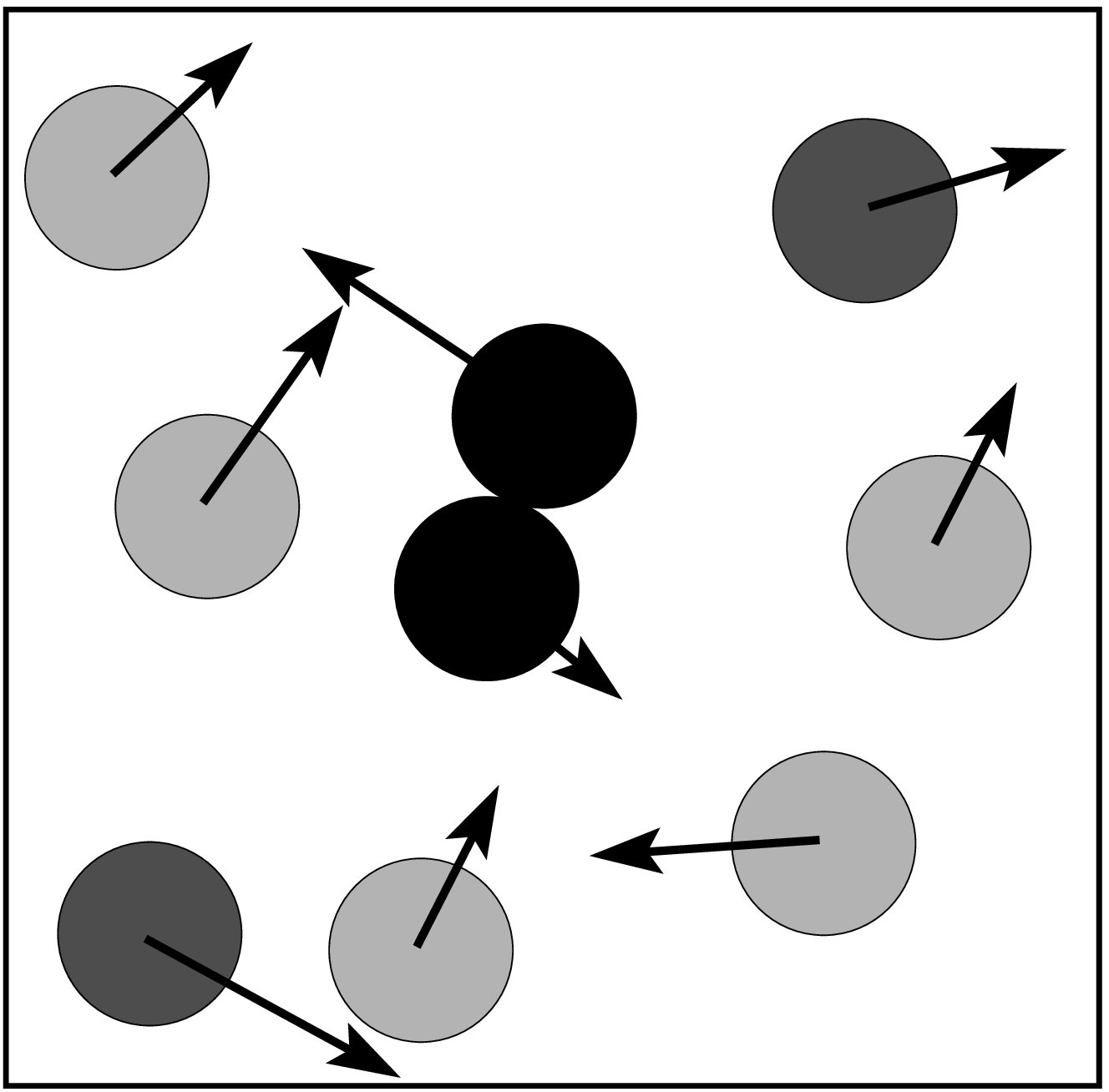,width=4cm}}
  \vspace*{13pt} \fcaption{The velocity shuffling procedure. Left: the
    black particles at the instant of the collision with pre-collision
    velocities. Middle: the colliding particles are assigned new
    velocities due to the collision rule Eq. (\ref{eq:coll}), two
    randomly selected particles are drawn dark grey. Right: the
    velocities of the colliding particles are shuffled with the
    velocities of the randomly selected particles. The shuffling
    procedure is unique due to the condition for the after-collision
    normal relative velocity to be positive.}
\label{fig:shuffle}
\end{figure}

Let us compare the analytical results for the temperature evolution
with the temperature decay as obtained by an event driven Molecular
Dynamics simulation. Haff's law was derived with the molecular chaos
assumption, therefore, in order to obtain results which are comparable
with theory we apply a special numerical procedure which destroys the
collision-induced velocity correlation of colliding particles. This
procedure is illustrated in Fig. \ref{fig:shuffle}. We shuffle the
velocities of the colliding particles with the velocities of randomly
selected particles. It is easy to understand that the shuffling
procedure is unique since the condition for the after-collision normal
relative velocity to be positive must be fulfilled.

\begin{figure}[htbp]
  \vspace*{13pt} \centerline{\psfig{file=Poeschel3a.eps,width=10cm}}
  \vspace{-4.8cm} \hspace{-1.8cm}
  \centerline{\psfig{file=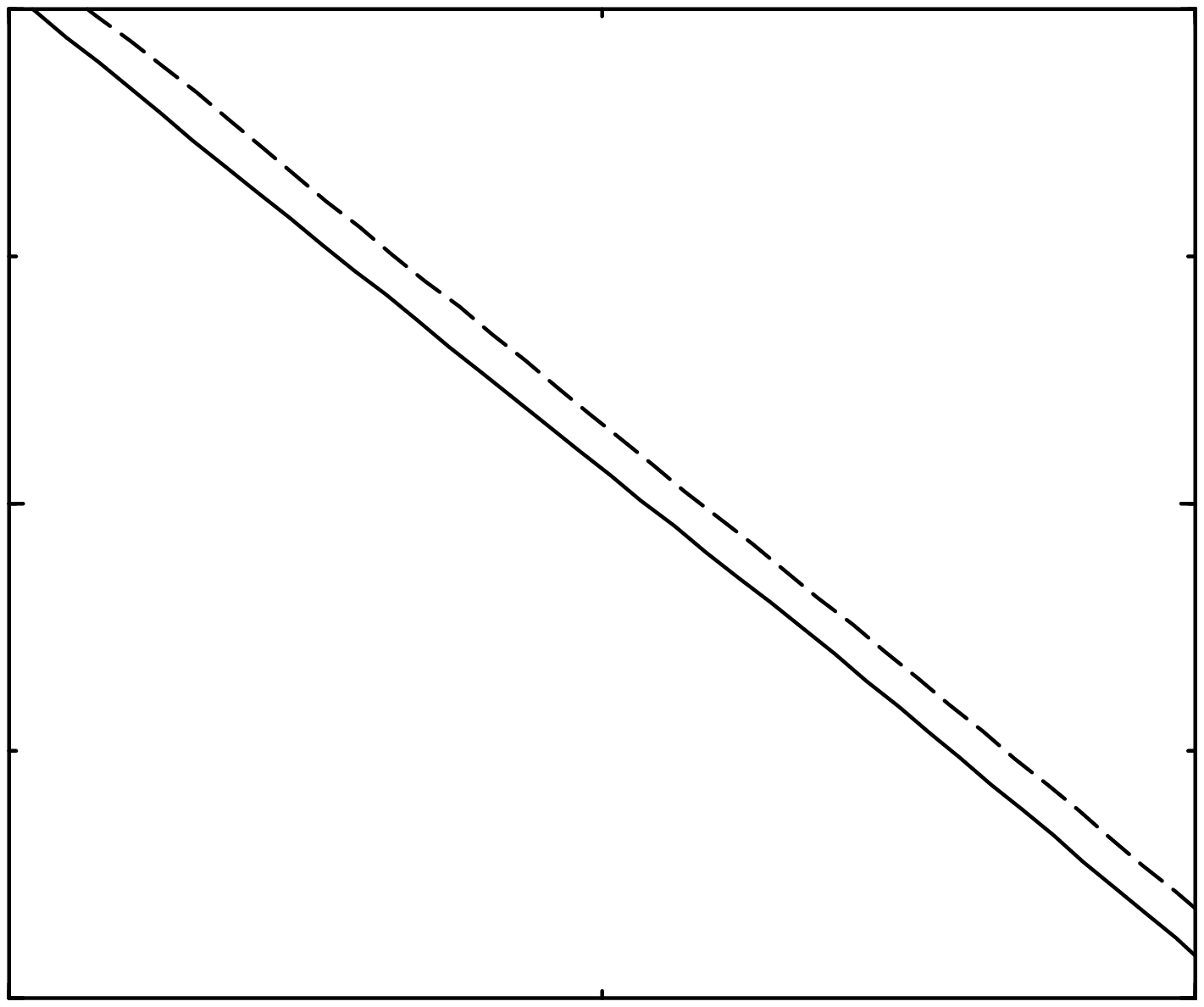,width=5cm}}
  \vspace{1.5cm} \fcaption{Evolution of the temperature. Dashed line:
    event driven MD simulation ($\epsilon=0.93$, $n=0.05$), full line:
    Haff's law, Eq. (\ref{eq:Haff}) together with the randomized event
    driven simulation (lines fall together). As soon as clusters
    appear ($\ln t\approx 6$) there is a large deviation of the
    numerical results from theory. The inset shows a magnification of
    the range $2.5 \le \ln t \le 3$. In the entire range of time we
    observe a small but systematic deviation between the curves.}
\label{fig:Tsimu}
\end{figure}

The solid line in Fig. \ref{fig:Tsimu} shows the evolution of
temperature for a system of $N=5,000$ inelastically colliding spheres
in two dimensions with periodic boundary conditions where the
described shuffling procedure has been applied (randomized
simulation). The reduced density was $n= N\sigma^2/V=0.05$
(with $\sigma=1$), where $V$ is the area of the quadratic box and the
coefficient of restitution is $\epsilon=0.93$. In the entire range of
$t$ the curve is identical up to the line width with Haff's law
(\ref{eq:Haff}) where the relaxation time $\tau_0$ is a fit parameter.
Repeating the same simulation without the shuffling procedure we
obtain a temperature decay as drawn by the dashed line in Fig.
\ref{fig:Tsimu}.  The curves are close to each other until $\ln
t\approx 6$. At this time instant for the unshuffled system we observe
formation of spatial inhomogeneities (clusters) which
leads to a reduced collision frequency. The cluster formation in
dissipative gases due to a pressure instability is a well understood
phenomenon which was described first in\cite{GoldhirschZanetti:1993}.

Obviously, our randomization procedure suppresses the formation of
density inhomogeneities which can be understood from mode-coupling
analysis.\cite{GoldhirschZanetti:1993} We know that the space
inhomogeneities arise in two steps: first vortex patterns appear due
to interparticle velocity correlations induced by the collisions, then
clusters develop owing to interaction between the density modes and
the flow-velocity modes. Thus, the collision-induced correlations play
a clue role in the formation of spatial inhomogeneities. Therefore,
since clustering or vertex formation is not possible in the randomized
simulation due to suppression of the collision-induced velocity
correlations we find almost perfect agreement with Haff's law.

Figure \ref{fig:Tsimu} shows that in the region $\ln t \lesssim 6$
Haff's law describes the temperature decay of the dissipative gas with
and without randomization up to a good degree of accuracy.  Having a
closer look to Fig. \ref{fig:Tsimu} (inset) one recognizes, however, a
small but systematic deviation between these two temperature curves.
We have checked spatial homogeneity of the particle distribution for
the both gases and did not observe any difference in the time span
where Haff's law is applicable. Thus we conclude that this difference
may be attributed only to velocity correlations of the particles of
the gas.

It is possible to fit the value of the relaxation time $\tau_0$ in
Haff's law to reproduce the simulation data (dashed line) up to the
same accuracy as done before for the randomized gas. The numerical
value of $\tau_0$ which depends on the coefficient of restitution and
on the density of the gas may, therefore, serve to characterize the
velocity correlations of the particles. We will return to this issue
in Sec. 6.

\section{Beyond the concept of molecular chaos}
\noindent 
To develop a theory beyond the concept of molecular chaos we introduce
the reduced velocities
\begin{equation}
  \label{eq:redvel}
  \vec{c}\equiv \frac{\vec{v}}{v_0} ~~~~\mbox{with}~~~~ 
v_0\equiv \sqrt{\frac{2T}{m}}\,,
\end{equation}
where $v_0$ is the thermal velocity and the reduced velocity
distribution function $\tilde{f}\left(\vec{c}\right)$ which is defined
by
\begin{equation}
  \label{eq:redf}
  f\left(\vec{v},t\right)=\frac{n}{v_0^3(t)}\tilde{f}\left(\vec{c},t\right)\,. 
\end{equation}
In the equivalent way we define the reduced two-particle distribution
function $\tilde{f}_2\left(\vec{c}_1,\vec{c}_2,\vec{r}_{12},t\right)$
with $\vec{r}_{12}\equiv \vec{r}_1-\vec{r}_2$.

Correlations of the particle velocities are taken into account by
assuming the following form of the two-particle velocity distribution
function:
\begin{equation}
  \label{eq:f2genform}
        \tilde{f}_2\left(\vec{c}_1,\vec{c}_2,\vec{r}_{12},t\right)= 
\tilde{f}\left(c_1\right)\tilde{f}\left(c_2\right) 
\left[g_2\left(r_{12}\right) 
+\chi\left(r_{12}\right) 
\hat{D}\left(\hat{\vec{r}}_{12}\right):\vec{c}_1\vec{c}_2 \right] \,
\end{equation} 
The first term in the brackets in the right-hand side of Eq.
(\ref{eq:f2genform}) accounts only for static correlations, while the
second term accounts for dynamic correlation of a pair of particles
with the reduced velocities $\vec{c}_1$ and $\vec{c}_2$. The traceless
dyad of the unit vector $\hat{\vec{r}}_{12}=\vec{r}_{12}/r_{12}$ is
defined as
\begin{equation}
  \label{eq:dyad_of_e}
        \hat{D}\left(\hat{\vec{r}}_{12}\right) \equiv 3 \, 
\hat{\vec{r}}_{12}\hat{\vec{r}}_{12} - \hat{U} \, 
\end{equation} 
with $\hat{U}$ being the unit tensor and we use the conventional
notation of vector algebra, i.e., $\vec{a}\vec{a}$ denotes the dyad
and
\begin{equation}
  \label{eq:c1Dc2}
  \hat{D}\left(\hat{\vec{r}}_{12}\right):\vec{c}_1\vec{c}_2 \equiv 3
\left(\vec{c}_1\cdot \hat{\vec{r}}_{12} \right)\left(\vec{c}_2\cdot
\hat{\vec{r}}_{12} \right)-\left(\vec{c}_1\cdot \vec{c}_2\right) \, .
\end{equation}
This is one of the simplest forms which may be built up from three
vectors and which features the requested symmetry with respect to
vectors $\vec{c}_1$ and $\vec{c}_2$. The function
$\chi\left(r_{12}\right)$ expresses the dependence of the velocity
correlations on the particle distance which, obviously, vanishes for
infinite distance, i.e., $\chi\left(r_{12}\right) \to 0$ for $r_{12}
\to \infty$.

At the instant of a collision, i.e., when $r_{12}=\sigma$ (or
$\hat{\vec{r}}_{12} =\vec{e}$) Eq. (\ref{eq:f2genform}) reads
\begin{equation}
  \label{eq:f2atsigma}
        \tilde{f}_2\left(\vec{c}_1,\vec{c}_2,\sigma \vec{e}\right)= 
\tilde{f}\left(c_1\right)\tilde{f}\left(c_2\right)g_2\left(\sigma\right)
\left\{ 1+b \left[3\left(\vec{c}_1 \cdot \vec{e}\right)\left(\vec{c}_2 \cdot 
\vec{e}\right)  - \left(\vec{c}_1 \cdot \vec{c}_2\right)\right] \right\}\, ,
\end{equation} 
where $b\equiv \chi\left(\sigma\right)/g_2\left(\sigma\right)$. For
the one-particle distribution function we use its representation in
terms of the Sonine polynomial expansion with the first non-vanishing
Sonine coefficient $a_2$:
\begin{equation}
  \label{eq:fSonin}
        \tilde{f}\left(\vec{c}_1\right)=\phi\left(c\right)
\left[1+a_2 S_2\left(c^2\right) \right] \, ,
\end{equation} 
where for the two-dimensional case the second Sonine polynomial reads, 
\begin{equation}
  \label{eq:Son2}
        S_2(x)=\frac12 x^2 -2 x + 1 \, .
\end{equation} 
With the distribution function given by Eqs.
(\ref{eq:f2atsigma},\ref{eq:fSonin},\ref{eq:Son2}) all necessary
quantities which describe the kinetic properties of the gas can be
obtained, e.g., the second moment of the collision integral
\begin{equation}
  \label{eq:defmu2}
\mu_2 =-\frac12 \int d\vec{c}_1\int d\vec{c}_2 \int d\vec{e} 
\Theta\left(-\vec{c}_{12} \cdot \vec{e}\right) \left|\vec{c}_{12} 
\cdot \vec{e}\right| 
\tilde{f}_2\left(\vec{c}_1,\vec{c}_2,\sigma \vec{e}\right) \Delta 
\left(c_1^2+c_2^2\right) \,, 
\end{equation} 
where $\Delta\left(c_1^2+c_2^2\right)$ denotes the change of the
quantity $c_1^2+c_2^2$ at the dissipative collision.

The importance of the second moment $\mu_2$ follows from the fact that
it characterizes the dissipation rate due to inelastic collisions.
Indeed, the factor $\Theta\left(-\vec{c}_{12} \cdot \vec{e}\right)$ in
the integrand guarantees that particles with velocities $\vec{c}_1$
and $\vec{c}_2$ do collide. The factor $\left|\vec{c}_{12} \cdot
  \vec{e}\right|$ characterizes the length of the collision cylinder,
which multiplied by the factor
$\tilde{f}_2\left(\vec{c}_1,\vec{c}_2,\sigma \vec{e}\right)$
characterizes the number of collisions between particles with
velocities $\vec{c}_1$, $\vec{c}_2$ and the intercenter vector $\sigma
\vec{e}$, occurring per unit time. The last factor in the integrand
$\Delta \left(c_1^2+c_2^2\right)$ characterizes the energy loss in
such collisions. Integration over all possible velocities and
intercenter vectors yields the quantity which characterizes the energy
loss per unit time, i.e., the decay rate of the gas temperature.

Straightforward calculations yield 
\begin{equation}
  \label{eq:mu2a2b}
\mu_2 = \frac12 \sqrt{2 \pi} \left(1-\epsilon^2\right)
\left[ 1+\frac{3}{16}a_2-\frac{63}{40} b\right] \, .
\end{equation} 
This expression shows that $\mu_2$ depends on both coefficients $a_2$
and $b$ in a similar way.

\section{Influence of the velocity correlations on the temperature evolution}
\label{sec:Influence}
\noindent Originally, Haff's law (\ref{eq:Haff}) was derived by scaling 
arguments with the assumption of the Maxwell distribution for the
particle velocities.\cite{Haff:83} From this derivation it is not
possible to obtain the relaxation time $\tau_0$ in terms of the
microscopic particle properties. Starting with the Boltzmann-Enskog
equation, however, the temperature
evolution\cite{BrilliantovPoeschelStability:2000}, i.e. Haff's law,
and expression for the relaxation time $\tau_0$ in terms of the second
moment of the collision integral can be found:
\begin{equation}
  \label{eq:tau0}
   \tau_0^{-1}=\frac13 g_2(\sigma) \sigma^2 n  \sqrt{\frac{2T_0}{m}}\, \, 
\mu_2 \, .  
\end{equation}
With expression (\ref{eq:mu2a2b}) for $\mu_2$ we obtain
\begin{equation}
  \label{eq:tau0_mu2}
        \tau_0=\frac{1}{1-\epsilon^2} A(\epsilon) \, , 
\end{equation}
with 
\begin{equation}
  \label{eq:A}
        A^{-1}(\epsilon)=\omega_0 \left[1+\frac{3}{16}a_2(\epsilon) 
-\frac{63}{40} b(\epsilon) \right] \,, ~~~~~~\omega_0=\frac13 g_2(\sigma) 
\sigma^2 n \sqrt{\pi \frac{T_0}{m}}\,.
\end{equation}
The second Sonine coefficient for a two-dimensional gas of inelastic
discs as a function of the coefficient of restitution has been
obtained with the molecular chaos assumption (i.e. ignoring particle
correlations)\cite{NoijeErnst:1998}:
\begin{equation}
\label{eq:a2linD}
        a_2 = \frac{16 (1-\epsilon ) \left(1 - 2 \epsilon^2\right)}
{ 57 -25 \epsilon +30 \epsilon^2
        ( 1 - \epsilon ) }\,.
\end{equation} 
Using Eqs. (\ref{eq:tau0_mu2},\ref{eq:A},\ref{eq:a2linD}) the
(macroscopic) relaxation time $\tau_0$ is expressed in terms of
microscopic quantities which characterize the particle collisions.

Analyzing Eqs. (\ref{eq:A},\ref{eq:a2linD}) it turns out that the
dependence of $A(\epsilon)$ on $\epsilon$ is very weak: In the
interval $\epsilon \in (0.8, 1.0)$ it varies within less than $0.6
\%$.  From Eq. (\ref{eq:A}) follows that the velocity correlations,
expressed by $b$ preserve the functional form of the temperature
decay, but change the value of the relaxation time. Moreover, it also
follows that the more pronounced the velocity correlations (i.e. the
larger the coefficient $b$) the more the relaxation time $\tau_0$
differs from the relaxation time of a gas where the velocity
correlations are suppressed. Taking into account that $a_2$ is small
and that $b$ is also not large, we can write for the factor
$A(\epsilon)$:
\begin{equation}
\label{eq:Afin}
        A=\omega_0^{-1} \left[1-\frac{3}{16}a_2(\epsilon) 
+\frac{63}{40} b(\epsilon) \right]  =
        A_{\rm rand} + \frac{63}{40} \omega_0^{-1} b(\epsilon) \, 
\end{equation} 
where $A_{\rm rand}$ is the factor $A$ that corresponds to a gas
without velocity correlations, i.e., a randomized gas.

\begin{figure}[htbp]
  \vspace*{13pt}
  \centerline{\psfig{file=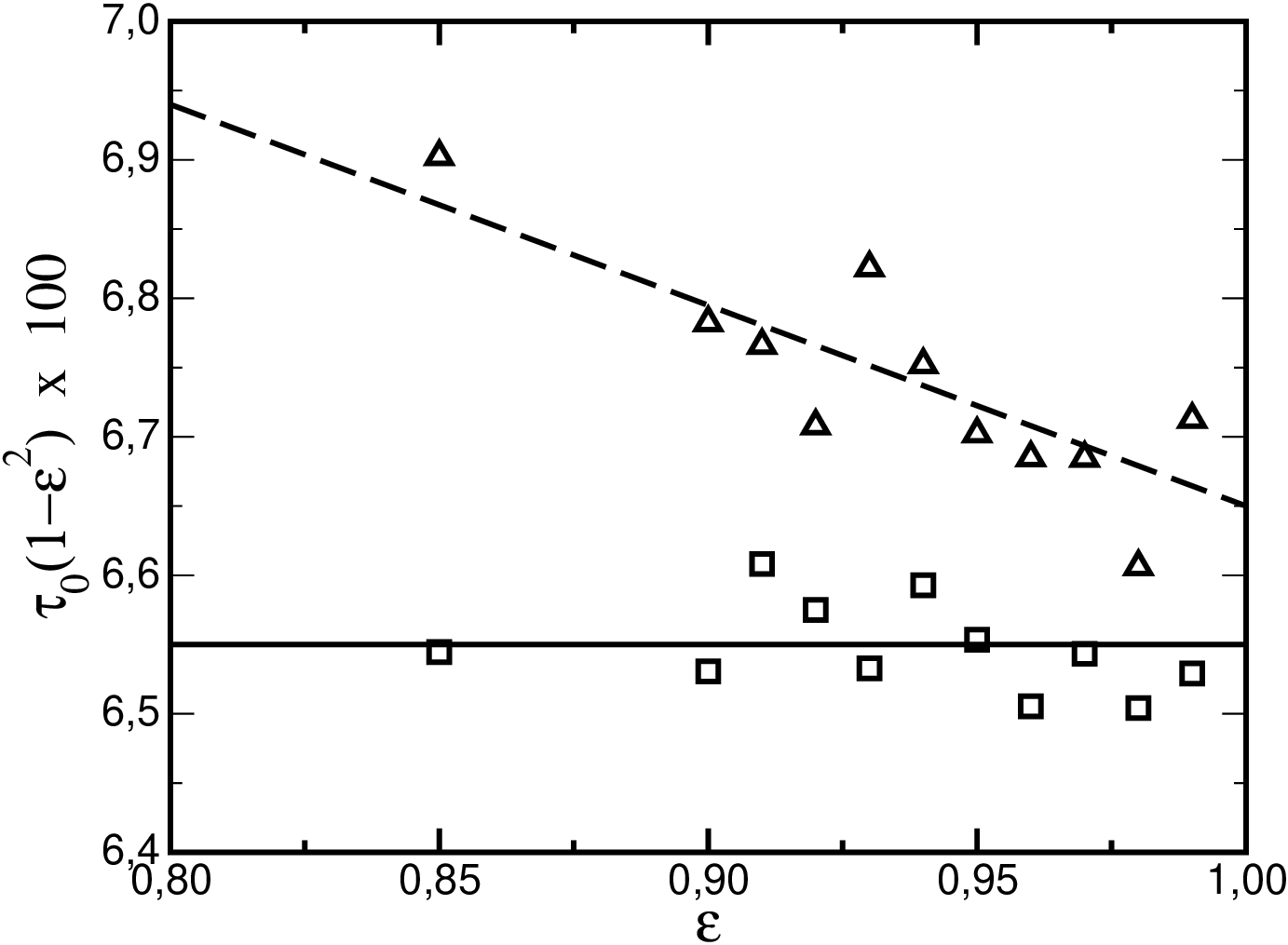,width=6.3cm,clip=}~~\psfig{file=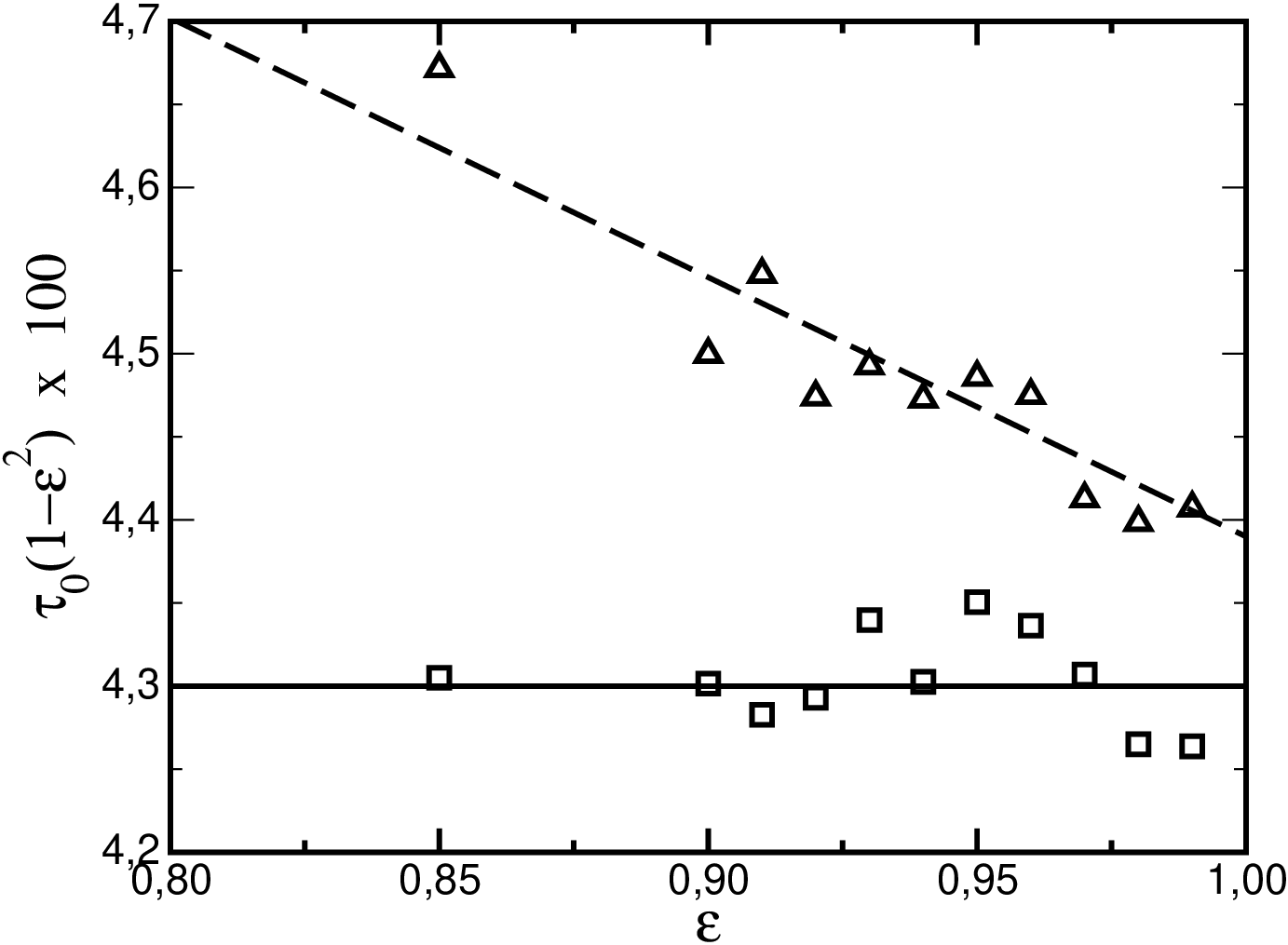,width=6.3cm,clip=}}
  \vspace*{13pt} 
\fcaption{The function $A(\epsilon)\equiv
    \tau_0\left(1-\epsilon^2\right)$ over the coefficient of
    restitution $\epsilon$ for a simulation of 5,000 inelastic hard
    disks of density $n=0.05$ (left) and $n=0.1$ (right). The lines
    have been fit to the numerical data. For the randomized simulation
    (squares) the curve does not significantly depend on $\epsilon$
    whereas in agreement with the theoretical prediction Eq.
    (\ref{eq:Afin}) the original gas shows noticeable dependence on
    $\epsilon$ due to velocity correlations.}
\label{fig:a05}
\end{figure}

This theoretical prediction can be checked by numerical simulations:
again we perform event driven Molecular Dynamics simulations, measure
the temperature as a function of time and determine the value of
$\tau_0$ as a fit parameter due to Eq. (\ref{eq:Haff}). According to
Eq. (\ref{eq:tau0_mu2}) the function $A(\epsilon)$ can be obtained by
drawing $\tau_0\left(1-\epsilon^2\right)$ over $\epsilon$. The results
are shown Fig. \ref{fig:a05} for two values of the density $n$.

The squares corresponds to the randomized gas and do not exhibit any
noticeable dependence on the coefficient of restitution $\epsilon$. As
it has been already mentioned this is a consequence of the very weak
dependence of $A$ on $\epsilon$ via the second Sonine coefficient
$a_2$. The triangles show the same data for simulations where the
randomizing procedure has not been applied. The lines show linear
least square fits to the numerical data.  Figure \ref{fig:a05} shows
that the dependence of $A$ on $\epsilon$ via the correlation
coefficient $b$ can be approximated by a linear function of $\epsilon$.

\section{Conclusions}
\noindent We study microscopic particle velocity correlations in 
dissipative gases. Taking into account velocity correlations preserves
the functional form of the temperature decay, which still obeys Haff's
law but changes the temperature relaxation time. Going beyond the
assumption of molecular chaos we suggest the functional form of the
two-particle distribution function which accounts for velocity
correlations and calculate the temperature relaxation time for this
distribution function. We performed Molecular Dynamics simulations of
dissipative gases and suggest a randomizing procedure which suppresses
the velocity correlations but leaves all other properties of the gas
dynamics unchanged. From the comparison of the results of the
simulation of the randomized gas with traditional simulations we
obtain the dependence of the correlation coefficient on the
coefficient of restitution.

\nonumsection{Acknowledgments}
\noindent
This research is supported by Deutsche Forschungsgemeinschaft via
grant (Po472/6).

\nonumsection{References}

\end{document}